\documentclass[aps,prb,twocolumn,showpacs]{revtex4}

\usepackage{graphicx}
\usepackage{amsmath}
\usepackage{amssymb}

\begin{document}

\title{Pseudogap-induced anisotropic suppression of electronic Raman response in cuprate superconductors}

\author{Pengfei Jing and Yiqun Liu}

\affiliation{Department of Physics, Beijing Normal University, Beijing 100875, China}

\author{Huaisong Zhao}

\affiliation{College of Physics, Qingdao University, Qingdao 266071, China}

\author{L\"ulin Kuang}

\affiliation{Sugon National Research Center for High-performance Computing Engineering Technology, Beijing 100093, China}

\author{Shiping Feng}
\email{spfeng@bnu.edu.cn}

\affiliation{Department of Physics, Beijing Normal University, Beijing 100875, China~~~}


\begin{abstract}
It has become clear that the anomalous properties of cuprate superconductors are intimately related to the formation of a pseudogap. Within the framework of the kinetic-energy-driven superconducting mechanism, the effect of the pseudogap on the electronic Raman response of cuprate superconductors in the superconducting-state is studied by taking into account the interplay between the superconducting gap and pseudogap. It is shown that the low-energy spectra almost rise as the cube of energy in the $B_{\rm 1g}$ channel and linearly with energy in the $B_{\rm 2g}$ channel. However, the pseudogap is strongly anisotropic in momentum space, where the magnitude of the pseudogap around the nodes is smaller than that around the antinodes, which leads to that the low-energy spectral weight of the $B_{\rm 1g} $ spectrum is suppressed heavily by the pseudogap, while the pseudogap has a more modest effect on the electronic Raman response in the $B_{\rm 2g}$ orientation.
\end{abstract}


\maketitle

It is now generally accepted that the unusual properties of cuprate superconductors in the underdoped and optimally doped regimes are heavily influenced by the pseudogap \cite{Hufner08,Basov05,Timusk99}. This pseudogap exists above the superconducing (SC) transition temperature $T_{\rm c}$ but below the pseudogap crossover temperature $T^{*}$, and manifests itself as a loss of the spectral weight of the low-energy quasiparticle excitations \cite{Hufner08,Basov05,Timusk99}. The origin of the pseudogap in cuprate superconductors remains a mystery second only to that of superconductivity itself, and it is widely believed that by the investigation of the former one might uncover essential insights for the understanding of the latter.

In order to fully understand the nature of the pseudogap state in cuprate superconductors, the two-particle Raman scattering experiments that can probe the quasiparticle dynamics on different regions of the Brillouin zone (BZ) have provided useful information for clarifying the nature of the pseudogap phase of cuprate superconductors \cite{Timusk99,Devereaux07}. In a striking contrast to the infrared measurements of the conductivity spectrum \cite{Basov05}, the electronic Raman scattering can be used to give directly two spectra $B_{\rm 1g}$ and $B_{\rm 2g}$, each representing a different average over the electron Fermi surface (EFS) \cite{Timusk99,Devereaux07}. After intensive investigations over more than two decades, it has been shown the electronic Raman response (ERR) spectrum in the SC-state is rather universal within the whole cuprate superconductors \cite{Timusk99,Devereaux07,Staufer92,Chen94,Venturini02,Hewitt02,Tacon06,Guyaid08,Blanc09,Prestel10,Loret16,Nemetschek97,Naeini00,Opel00,Gallais05,Sakai13}, where the broad continuum scattering is depressed at low energies and the spectral weight lost is transferred to higher energies, forming a broad peak whose position depends on the scattering geometry. In particular, the earliest evidence of the pseudogap effect manifested itself as a suppression of the SC-state ERR around the antinodal regime was observed on YBa$_{2}$Cu$_{3}$O$_{7-x}$ in the underdoped regime \cite{Nemetschek97}. Later, these early observations have been further confirmed \cite{Naeini00,Opel00,Gallais05,Sakai13}, where a heavy loss in the spectral weight was found in the $B_{\rm 1g}$ spectrum, and then the overall low-energy spectral weight in the $B_{\rm 1g}$ spectrum is suppressed heavily by the pseudogap, however, only a light loss of the spectral weight was detected in the $B_{\rm 2g}$ spectrum, indicating a strong momentum dependence of the pseudogap in cuprate superconductors. In this case, the electronic Raman scattering therefore gives the important information for the nature of the momentum dependence of the pseudogap which is not accessible via the conductivity measurement.

Theoretically, ERR of cuprate superconductors in the SC-state has been studied extensively \cite{Timusk99,Devereaux07,Geng10,Muschler10,Chubukov06,Manske97,Branch95,Devereaux94,Greco14}. In particular, the ERR spectra in both the $B_{\rm 1g}$ and $B_{\rm 2g}$ orientations have been extracted to fit the electron-boson spectral density, and the results provide information on the angular variation of the electron self-energy and the corresponding spectral density around EFS \cite{Muschler10}. Moreover, it has been shown that ERR in the $B_{\rm 1g}$ symmetry allows one to distinguish between phonon-mediated and magnetically mediated d-wave superconductivity \cite{Chubukov06}. Furthermore, the main features observed in ERR have been analysed by the self-energy effect in the proximity to a d-wave flux-phase order instability \cite{Greco14}. However, to the best of our knowledge, the anisotropic suppression of the ERR spectrum by the momentum dependence of the pseudogap in cuprate superconductors has not been treated starting from a microscopic theory. In this paper, we study the pseudogap-induced anisotropic suppression of the ERR spectrum in cuprate superconductors in the SC-state by taking into account the interplay between the pseudogap and SC gap. Within the framework of the kinetic-energy-driven SC mechanism \cite{Feng0306,Feng12,Feng15}, we calculate the ERR function in terms of the Raman density-density correlation function, and then the obtained results show that the low-energy $B_{\rm 2g}$ response depends linearly on energy $\omega$, while the low-energy $B_{\rm 1g}$ spectrum varies as $\omega^{3}$. However, our results also show that the pseudogap on EFS is strong momentum dependent, with the actual minimum that does not appear around the nodes, but locates exactly at the hot spots. In particular, the magnitude of the pseudogap around the nodes is smaller than that around the antinodes. This special structure of the momentum dependence of the pseudogap therefore leads to that the low-energy spectral weight of the $B_{\rm 1g}$ spectrum is suppressed heavily by the pseudogap, while the pseudogap has a more modest effect on the ERR spectrum in the $B_{\rm 2g}$ orientation.


ERR is manifested itself by the dynamical Raman response function $\tilde{S}(\omega)$, which can be obtained directly from the imaginary part of the Raman density-density correlation function $\tilde{\chi}({\bf q},\omega)$ as \cite{Devereaux07},
\begin{eqnarray}\label{ERRF}
\tilde{S}(\omega)=-{1\over\pi}[1+n_{B}(\omega)]{\rm Im}\tilde{\chi}({{\bf q}\sim 0},\omega),
\end{eqnarray}
where $n_{B}(\omega)$ is the boson distribution function, while the Raman density-density correlation function $\tilde{\chi}({\bf q},\omega)$ is defined as,
\begin{eqnarray}\label{RDDCF}
\tilde{\chi}({\bf q},\tau-\tau')=-\langle T\rho_{\gamma}({\bf q},\tau) \rho_{\gamma}(-{\bf q},\tau')\rangle,
\end{eqnarray}
where the Raman density operator in the Nambu representation can be expressed as,
\begin{eqnarray}\label{RDO}
\rho_{\gamma}({\bf q})=\sum_{{\bf k}}\gamma^{\alpha}_{{\bf k}+{{\bf q}\over 2}}C^{\dagger}_{{\bf k}+{\bf q}}\tau_{3}C_{\bf k},
\end{eqnarray}
with the Pauli matrix $\tau_{3}$, and the bare Raman vertex $\gamma^{\alpha}_{\bf k}$ that has been classified by the representations $B_{\rm 1g}$ and $B_{\rm 2g}$ of the point group $D_{\rm 4h}$ as \cite{Devereaux07},
\begin{subequations}\label{Vertex}
\begin{eqnarray}
\gamma^{B_{\rm 1g}}_{\bf k}&=& {1\over 4}b_{\omega_{i},\omega_{s}}[\cos(k_{x})-\cos(k_{y})],\\
\gamma^{B_{\rm 2g}}_{\bf k}&=& b'_{\omega_{i},\omega_{s}}\sin(k_{x})\sin(k_{y}),
\end{eqnarray}
\end{subequations}
respectively, where as a qualitative discussion, the magnitude of the energy dependence of the prefactors $b$ and $b'$ can be rescaled to units. In cuprate superconductors, the $B_{\rm 1g}$ spectrum samples the antinodal region, the Fermi momentum on the BZ boundary, while the $B_{\rm 2g}$ spectrum samples the nodal region, therefore ERR probes complementary regimes of EFS \cite{Devereaux07}. Substituting the Raman density operator (\ref{RDO}) into Eq. (\ref{RDDCF}), the Raman density-density correlation function $\tilde{\chi}({\bf q},\omega)$ therefore can be rewritten in terms of the single-electron Green's function $\mathbb{G}({\bf k},\omega)$ in the Nambu representation as,
\begin{eqnarray}\label{RDDCF1}
&~&\tilde{\chi}({\bf q},iq_{m})={1\over N}\sum_{\bf k}\gamma^{\alpha}_{1{\bf k}+{1\over 2}{\bf q}}\gamma^{\alpha}_{2{\bf k}+{1\over 2}{\bf q}} \nonumber\\
&\times& {1\over \beta} \sum_{i\omega_{n}}{\rm Tr}[\mathbb{G}({\bf k}+{\bf q},i\omega_{n}+iq_{m})\tau_{3}\mathbb{G}({\bf k},i\omega_{n})\tau_{3}].
\end{eqnarray}
It should be emphasized that in the obtaining above Raman density-density correlation function, the vertex correction has been ignored, since it has been shown that the vertex correction in the pseudogap phase is negligibly small \cite{Lin12,Bergeron11}. Eq. (\ref{RDDCF1}) also indicates that for the evaluation of the electronic Raman density-density correlation function in the SC-state, we firstly need to obtain the single-electron Green's function.

The $t$-$J$ model on a square-lattice is widely accepted for the description of the essential physics of cuprate superconductors \cite{Anderson87,Phillips10}. This $t$-$J$ model is defined as,
\begin{eqnarray}\label{tjham}
H&=&-\sum_{ll'\sigma}t_{ll'}C_{l\sigma}^{\dagger}C_{l'\sigma}+\mu\sum_{l\sigma}C_{l\sigma}^{\dagger}C_{l\sigma}+J\sum_{<ll'>}S_{l}\cdot S_{l'},~~~~
\end{eqnarray}
supplemented by an important on-site local constraint to avoid the double occupancy, $\sum_{\sigma}C^{\dagger}_{l\sigma} C_{l\sigma}\leq 1$, where $C_{l\sigma}^{\dagger}$ and $C_{l\sigma}$ are the electron operators that respectively create and annihilate electrons with spin $\sigma$ on the lattice site $l$, ${\bf S}_{l}$ is a spin operator located on the lattice site $l$, $\mu$ is the chemical potential, $J$ is the magnetic interaction between the nearest-neighbor (NN) sites on a square-lattice, and the summation $<ll'>$ is carried over the NN bonds. For the hopping $t_{ll'}$, we take only $t$ for the NN hopping amplitude and $-t'$ for the next NN hopping amplitude, respectively. In this paper, these parameters are chosen as $t/J=2.5$ and $t'/t=0.3$. The effect of the strong electron correlation in the $t$-$J$ model (\ref{tjham}) manifests itself by the no-double electron occupancy local constraint \cite{Lee06,Feng93,Zhang93,Yu92,Anderson00}. It has been shown that this no-double electron occupancy local constraint can be treated properly in actual calculations within the fermion-spin theory \cite{Feng15,Feng9404}, where the constrained electron is decoupled as a charge carrier and a localized spin, i.e., the electron operators $C_{l\uparrow}$ and $C_{l\downarrow}$ are decoupled as $C_{l\uparrow}=h^{\dagger}_{l\uparrow}S^{-}_{l}$ and $C_{l\downarrow}=h^{\dagger}_{l\downarrow}S^{+}_{l}$, respectively, with the charge carrier $h_{l\sigma}=e^{-i\Phi_{l\sigma}}h_{l}$ that represents the charge degree of freedom together with some effects of spin configuration rearrangements due to the presence of the doped charge carrier itself, while the localized spin $S_{l}$ describes the spin degree of freedom. In this fermion-spin representation, the original $t$-$J$ model (\ref{tjham}) can be rewritten as,
\begin{eqnarray}\label{cssham}
H&=&\sum_{ll'}t_{ll'}(h^{\dagger}_{l\uparrow}h_{l'\uparrow}S^{-}_{l}S^{+}_{l'}+h^{\dagger}_{l\downarrow}h_{l'\downarrow}S^{+}_{l}S^{-}_{l'}) \nonumber\\
&-& \mu \sum_{l\sigma}h^{\dagger}_{l\sigma} h_{l\sigma}+J_{{\rm eff}}\sum_{\langle ll'\rangle}{\bf S}_{l}\cdot {\bf S}_{l'},~~~~
\end{eqnarray}
Following this $t$-$J$ model in the fermion-spin representation (\ref{cssham}), the kinetic-energy-driven SC mechanism has been developed \cite{Feng15,Feng0306,Feng12}, where the interaction between the charge carriers and spin directly from the kinetic energy by the exchange of spin excitations generates the SC-state in the particle-particle channel and pseudogap state in the particle-hole channel, therefore there is an interplay between the SC gap and pseudogap in the whole SC dome \cite{Feng12}.

In the framework of the charge-spin separation, the single-electron Green's function is obtained in terms of the charge-spin recombination \cite{Lee06,Feng93,Zhang93,Yu92,Anderson00}. However, how a microscopic theory based on the charge-spin separation that can give a consistent description of EFS in terms of the charge-spin recombination is a very difficult problem \cite{Feng93,Zhang93}. Recently, we \cite{Feng15a} have developed a full charge-spin recombination scheme to fully recombine a charge carrier and a localized spin into a constrained electron, where the obtained electron propagator can give a consistent description of the nature of EFS in cuprate superconductors. In the following discussions, we reproduce only the main details in the calculations of the single-electron Green's function of the $t$-$J$ model under the fermion-spin representation in the SC-state. In Ref. \onlinecite{Feng15a}, the single-electron diagonal and off-diagonal Green's functions of the $t$-$J$ model in the SC-sate have been obtained in terms of the full charge-spin recombination scheme as,
\begin{widetext}
\begin{subequations}\label{SEGFS}
\begin{eqnarray}
G({\bf k},\omega)&=&{\omega+\varepsilon_{\bf k}+\Sigma_{1}({\bf k},-\omega)\over[\omega-\varepsilon_{\bf k}-\Sigma_{1}({\bf k},\omega)][\omega+ \varepsilon_{\bf k}+\Sigma_{1}({\bf k},-\omega)]-\bar{\Delta}^{2}({\bf k})}, ~~~\label{DSEGF}\\
\Im^{\dagger}({\bf k},\omega)&=&-{\bar{\Delta}({\bf k})\over [\omega-\varepsilon_{\bf k}-\Sigma_{1}({\bf k},\omega)][\omega+\varepsilon_{\bf k} +\Sigma_{1}({\bf k},-\omega)]-\bar{\Delta}^{2}({\bf k})},~~~\label{ODSEGFS}
\end{eqnarray}
\end{subequations}
\end{widetext}
where the bare electron excitation spectrum $\varepsilon_{\bf k}=-Zt\gamma_{\bf k}+Zt'\gamma_{\bf k}'+\mu$, with $\gamma_{\bf k}=({\rm cos}k_{x}+{\rm cos}k_{y})/2$, $\gamma_{\bf k}'= {\rm cos} k_{x}{\rm cos}k_{y}$, $Z$ is the number of the NN or next NN sites on a square lattice, and the d-wave SC gap $\bar{\Delta}({\bf k})=\Sigma_{2}({\bf k},\omega=0)=\bar{\Delta}\gamma^{(\rm d)}_{{\bf k}}$, with $\gamma^{(\rm d)}_{\bf k}=({\rm cos} k_{x}-{\rm cos} k_{y})/2$, while the electron self-energies $\Sigma_{1}({\bf k},\omega)$ in the particle-hole channel and $\Sigma_{2}({\bf k}, \omega)$ in the particle-particle channel due to the interaction between electrons by the exchange of spin excitations are evaluated in terms of the spin bubble, and have been given explicitly in Ref. \onlinecite{Feng15a}.

In the previous discussions \cite{Feng15a}, we on the other hand have shown that the momentum dependence of the pseudogap is directly related to the electron self-energy $\Sigma_{1}({\bf k},\omega)$ in the particle-hole channel as,
\begin{eqnarray}\label{PG}
\Sigma_{1}({\bf k},\omega)\approx {[\bar{\Delta}_{\rm PG}({\bf k})]^{2}\over\omega+\varepsilon_{0{\bf k}}},
\end{eqnarray}
where $\varepsilon_{0{\bf k}}=L^{({\rm e})}_{2}({\bf k})/L^{({\rm e})}_{1}({\bf k})$, $\bar{\Delta}_{\rm PG}({\bf k})$ is so-called as the pseudogap, and can be obtained as, $\bar{\Delta}_{\rm PG}({\bf k})=L^{({\rm e})}_{2}({\bf k})/\sqrt{L^{({\rm e})}_{1}({\bf k})}$, while $L^{({\rm e})}_{1}({\bf k})=-\Sigma_{\rm 1o}({\bf k}, \omega=0)$ and $L^{({\rm e} )}_{2}({\bf k}) =\Sigma_{1}({\bf k},\omega=0)$ are evaluated directly from $\Sigma_{1}({\bf k},\omega)$ given explicitly in Ref. \onlinecite{Feng15a}.

With the help of the self-energy $\Sigma_{1}({\bf k},\omega)$ in Eq. (\ref{PG}), we therefore obtain the single-electron diagonal and off-diagonal Green's functions in Eq. (\ref{SEGFS}) explicitly as,
\begin{subequations}\label{SEGFSPG}
\begin{eqnarray}
G({\bf k},\omega)&=&\sum_{\nu=1,2}\left ({U^{2}_{\nu{\bf k}}\over\omega-E_{\nu{\bf k}}}+{V^{2}_{\nu{\bf k}}\over \omega+ E_{\nu{\bf k}}}\right ), \label{DSEGFSPG}\\
\Im^{\dagger}({\bf k},\omega)&=&\sum_{\nu=1,2}{a_{\nu{\bf k}}\bar{\Delta}({\bf k})\over 2E_{\nu{\bf k}}}\left ({1\over\omega+E_{\nu{\bf k}}}- {1\over \omega-E_{\nu{\bf k}}}\right ), ~~~~~~~ \label{ODSEGFSPG}
\end{eqnarray}
\end{subequations}
where $a_{1{\bf k}}=(E^{2}_{1{\bf k}}-\varepsilon^{2}_{0{\bf k}})/(E^{2}_{1{\bf k}}-E^{2}_{2{\bf k}})$, $a_{2{\bf k}}= (E^{2}_{2{\bf k}}-\varepsilon^{2}_{0{\bf k}})/(E^{2}_{1{\bf k}}-E^{2}_{2{\bf k}})$, however, the SC quasiparticle spectrum has been divided into two branches, $E_{1{\bf k}}=\sqrt{[\Phi_{1{\bf k}}+\Phi_{2{\bf k}}]/2}$ and $E_{2{\bf k}}=\sqrt{[\Phi_{1{\bf k}}-\Phi_{2{\bf k}}]/2}$, respectively, by the pseudogap, where $\Phi_{1{\bf k}}=\varepsilon^{2}_{\bf k}+\varepsilon^{2}_{0{\bf k}}+ 2 \bar{\Delta}^{2}_{\rm PG}({\bf k})+ \bar{\Delta}^{2}({\bf k})$, $\Phi_{2{\bf k}}=\sqrt{(\varepsilon^{2}_{\bf k}-\varepsilon^{2}_{0{\bf k}})b_{1{\bf k} }+4\bar{\Delta}^{2}_{\rm PG}({\bf k})b_{2{\bf k}}+\bar{\Delta}^{4}({\bf k})}$, $b_{1{\bf k}}=\varepsilon^{2}_{\bf k}- \varepsilon^{2}_{0{\bf k}} +2\bar{\Delta}^{2}({\bf k})$, and $b_{2{\bf k}}= (\varepsilon_{\bf k}-\varepsilon_{0{\bf k}})^{2}+\bar{\Delta}^{2}({\bf k})$, while the coherence factors, $U^{2}_{1{\bf k}}= [a_{1{\bf k}}(1+\varepsilon_{\bf k}/E_{1{\bf k}})-a_{3{\bf k}}(1+\varepsilon_{0{\bf k}}/E_{1{\bf k}})]/2$, $V^{2}_{1{\bf k}}= [a_{1{\bf k}}(1-\varepsilon_{\bf k}/E_{1{\bf k}})-a_{3{\bf k}}(1-\varepsilon_{0{\bf k}}/E_{1{\bf k}})]/2$, $U^{2}_{2{\bf k}}=- [a_{2{\bf k}}(1+\varepsilon_{\bf k}/E_{2{\bf k}})-a_{3{\bf k}}(1+\varepsilon_{0{\bf k}}/E_{2{\bf k}})]/2$, and $V^{2}_{2{\bf k} }=- [a_{2{\bf k}}(1-\varepsilon_{\bf k}/E_{2{\bf k}})-a_{3{\bf k}}(1-\varepsilon_{0{\bf k}}/E_{2{\bf k}})]/2$, with $a_{3{\bf k}}=\bar{\Delta}^{2}_{\rm PG}({\bf k})/(E^{2}_{1{\bf k}}-E^{2}_{2{\bf k}})$.

Substituting the Green's function (\ref{SEGFSPG}) into Eqs. (\ref{RDDCF1}) and (\ref{ERRF}), the ERR function $\tilde{S}(\omega)$ is therefore obtained explicitly as,
\begin{eqnarray}\label{ERRF1}
\tilde{S}(\omega)&=&[1+n_{B}(\omega)]{2\over N}\sum_{{\bf k}\mu\nu}\gamma^{\alpha}_{1{\bf k}}\gamma^{\alpha}_{2{\bf k}}\{L^{(1)}_{\mu\nu}({\bf k}) \nonumber\\
&\times& [A^{(1)}_{\mu\nu} ({\bf k})\delta(\omega+E_{\nu{\bf k}}-E_{\mu{\bf k}})\nonumber\\ 
&-&A^{(2)}_{\mu\nu}({\bf k})\delta(\omega-E_{\nu{\bf k}}+ E_{\mu{\bf k}})]\nonumber\\
&+& L^{(2)}_{\mu\nu}({\bf k})[A^{(3)}_{\mu\nu}({\bf k})\delta(\omega-E_{\nu{\bf k}}-E_{\mu{\bf k}})\nonumber\\
&-&A^{(4)}_{\mu\nu}({\bf k})\delta(\omega+E_{\nu{\bf k}}+E_{\mu{\bf k}})]\},
\end{eqnarray}
where $L^{(1)}_{\mu\nu}({\bf k})=n_{\rm F}(E_{\nu{\bf k}})-n_{\rm F}(E_{\mu{\bf k}})$, $L^{(2)}_{\mu\nu}({\bf k})=1-n_{\rm F}(E_{\nu{\bf k}})-n_{\rm F}(E_{\mu{\bf k}})$, $A^{(1)}_{\mu\nu}({\bf k})=U^{2}_{\mu{\bf k}}U^{2}_{\nu{\bf k}}-\Xi_{\mu\nu}({\bf k})$, $A^{(2)}_{\mu\nu}({\bf k})= V^{2}_{\mu{\bf k}}V^{2}_{\nu{\bf k}}-\Xi_{\mu\nu}({\bf k})$, $A^{(3)}_{\mu\nu}({\bf k})=U^{2}_{\mu{\bf k}}V^{2}_{\nu{\bf k}}-\Xi_{\mu\nu}({\bf k})$, $A^{(4)}_{\mu\nu}({\bf k})=V^{2}_{\mu{\bf k}}U^{2}_{\nu{\bf k}}-\Xi_{\mu\nu}({\bf k})$, and $\Xi_{\mu\nu}({\bf k})=a_{\mu{\bf k} }a_{\nu{\bf k}} \bar{\Delta}^{2}({\bf k})/(4E_{\mu{\bf k}}E_{\nu{\bf k}})$.

\begin{figure}[h!]
\centering
\includegraphics[scale=0.3]{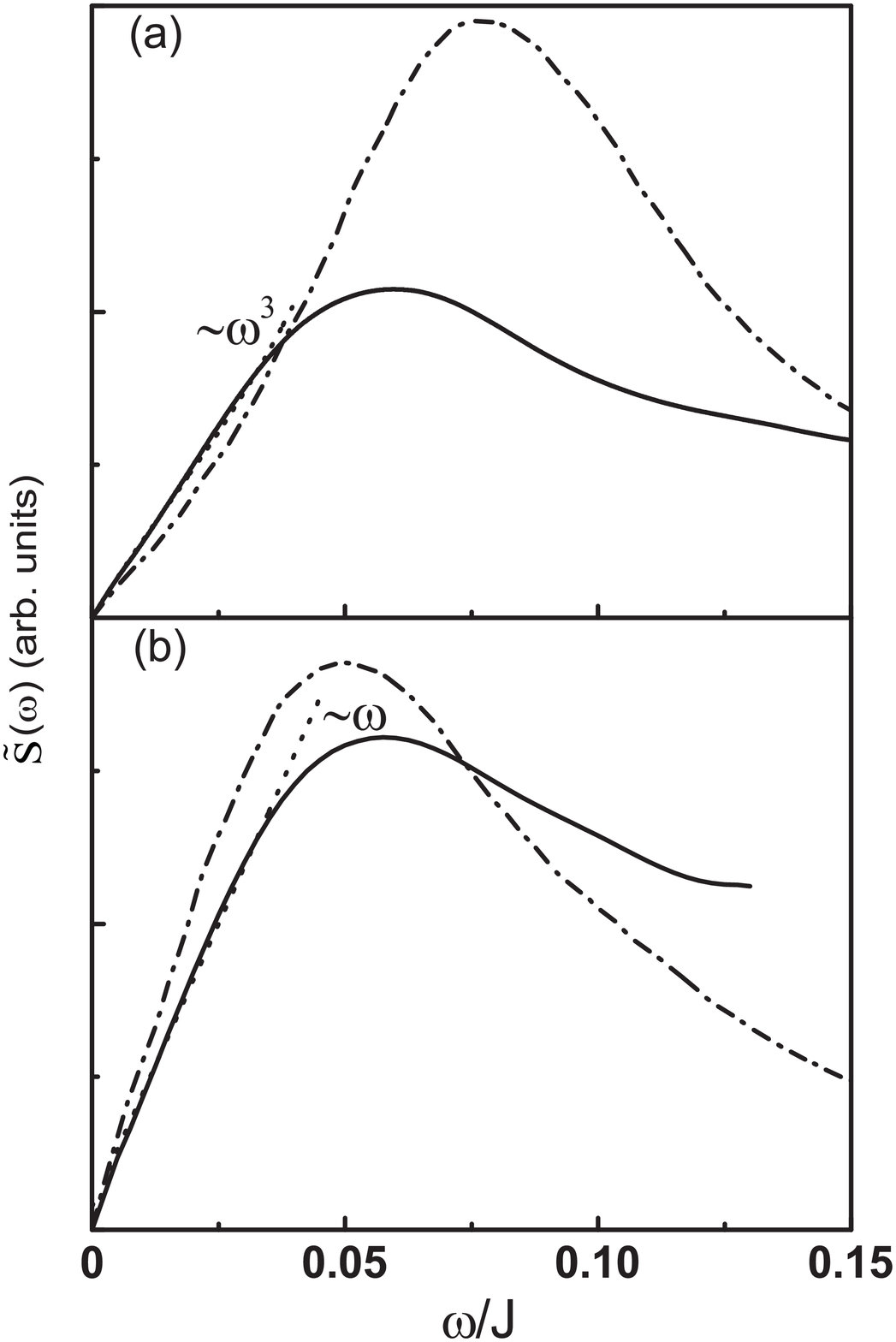}
\caption{(a) $B_{\rm 1g}$ and (b) $B_{\rm 2g}$ spectra (solid line) as a function of energy at $\delta=0.15$ with $T=0.002J$ for $t/J=2.5$ and $t'/t=0.3$. The dotted lines are a cubic and a linear fits for the low-energy $B_{\rm 1g}$ and $B_{\rm 2g}$ spectra, respectively. For comparison, the corresponding results of the $B_{\rm 1g}$ and $B_{\rm 2g}$ spectra in the case of the absence of the pseudogap (dash-dotted lines) are also shown in (a) and (b), respectively. \label{B1gB2g}}
\end{figure}

We are now ready to discuss ERR of cuprate superconductors in the SC-state. In the following discussions, we only focus on the low-energy features related to the anisotropic suppression of the low-energy spectral weight by the momentum dependence of the pseudogap. As a comparison of the results between the cases of the presence and absence of the pseudogap, we firstly discuss the case of the absence of the pseudogap, i.e., $\bar{\Delta}_{\rm PG}=0$. In this case, the ERR function $\tilde{S}(\omega)$ in Eq. (\ref{ERRF1}) is reduced as,
\begin{eqnarray}\label{ERRF2}
&~&\tilde{S}^{(0)}({\bf q},\omega)=[1+n_{B}(\omega)]{1\over 2N}\sum_{\bf k}\gamma^{\alpha}_{1{\bf k}}\gamma^{\alpha}_{2{\bf k}} {\bar{\Delta}^{2}({\bf k} ) \over E^{(0)2}_{\bf k}}\nonumber\\
&\times& {\rm th}\left ({1\over 2}\beta E^{(0)}_{\bf k}\right )[\delta(\omega-2E^{(0)}_{\bf k})-\delta(\omega+2E^{(0)}_{\bf k} )],~~~~
\end{eqnarray}
with $E^{(0)}_{\bf k}= \sqrt{\varepsilon^{2}_{\bf k}+\mid\bar{\Delta}({\bf k})\mid^{2}}$. We have performed a calculation for the ERR function (\ref{ERRF2}) in both $B_{\rm 1g}$ and $B_{\rm 2g}$ orientations, and the results of the $B_{\rm 1g}$ and $B_{\rm 2g}$ spectra (dash-dotted line) at doping $\delta=0.15$ with temperature $T=0.002J$ are plotted in Fig. \ref{B1gB2g}a and Fig. \ref{B1gB2g}b, respectively, where both $B_{\rm 1g}$ and $B_{\rm 2g}$ spectra are characterized clearly by the presence of the pair-breaking peaks, however, the peak in the $B_{\rm 1g}$ spectrum is located around the energy $\omega=2\bar{\Delta}$, while the intensity of the $B_{\rm 2g}$ spectrum is weaker than the $B_{\rm 1g}$ one and the peak position is located at a lower energy \cite{Timusk99,Devereaux07,Staufer92,Chen94,Venturini02,Hewitt02,Tacon06,Guyaid08,Blanc09,Prestel10,Loret16}.

However, when the pseudogap effect is included in terms of the electron self-energy $\Sigma_{1}({\bf k},\omega)$, the spectral weight of the ERR spectrum is suppressed. To see this point clearly, we also plot the results of the $B_{\rm 1g}$ and $B_{\rm 2g}$ spectra (solid line) in case of the presence of the pseudogap $\bar{\Delta}_{\rm PG}$ obtained from the Eq. (\ref{ERRF1}) at $\delta=0.15$ with $T=0.002J$ in Fig. \ref{B1gB2g}a and Fig. \ref{B1gB2g}b, respectively. In comparison with the corresponding results in the absence of the pseudogap (dash-dotted line), we therefore find that although the low-energy spectral weight of the ERR spectrum in the both $B_{\rm 1g}$ and $B_{\rm 2g}$ symmetries has been redistributed, the low-energy spectral weight in the $B_{\rm 1g}$ channel is suppressed heavily by the pseudogap, while the spectral weight in the $B_{\rm 2g}$ channel is reduced lightly, in qualitative agreement with the experimental data \cite{Nemetschek97,Naeini00,Opel00,Gallais05,Sakai13}. In particular, we have also calculated the energy dependence of the ERR functions up to higher energies, and the results show that the redistribution of the spectral weight induced by the pseudogap leads to a transfer of the missing low-energy spectral weight to the higher-energy region \cite{Hewitt02,Nemetschek97,Naeini00,Opel00,Gallais05,Sakai13}. Moreover, a numerical fit has been made to the low-energy data, and the results show that in the depleted low-energy region of the $B_{\rm 2g}$ spectrum, the intensity rises linearly with energy $\omega$, while the low-energy spectrum varies as $\omega^{3}$ in the $B_{\rm 1g}$ channel (see the dotted line in Fig. \ref{B1gB2g}), which are also consistent with the experimental data \cite{Timusk99,Devereaux07,Staufer92,Chen94,Venturini02,Hewitt02,Tacon06,Guyaid08,Blanc09,Prestel10,Loret16,Nemetschek97,Naeini00,Opel00,Gallais05,Sakai13}. On the other hand, these results are in a striking contrast to the conventional superconductors, where the ERR spectrum in the low-energy regime for an isotropic s-wave gap is characterized by the exponentially activated behavior. Furthermore, we have made a series of calculations for the $B_{\rm 2g}$ spectrum at different doping levels, and the results show that the $B_{\rm 2g}$ spectrum has a domelike shape of the doping dependence, and actually scales with $T_{\rm c}$ throughout the doping range.

\begin{figure}[h!]
\centering
\includegraphics[scale=0.35]{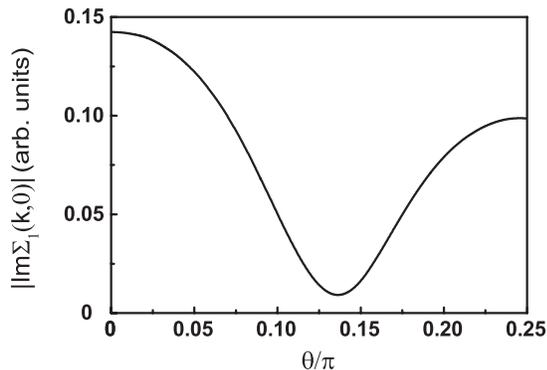}
\caption{The angular dependence of the pseudogap on the electron Fermi surface at $\delta=0.15$ with $T=0.002J$ for $t/J=2.5$ and $t'/t=0.3$. \label{pseudogap}}
\end{figure}

The poles of the electron Green's function (\ref{DSEGF}) at zero energy determine directly EFS in momentum space, and the everything on the other hand happens at EFS. An explanation of the anisotropic suppressions of ERR in cuprate superconductors in the SC-state can be found from the electron self-energy $\Sigma_{1}({\bf k},\omega)$ in Eq. (\ref{PG}) in the particle-hole channel, where the momentum dependence of the pseudogap is related explicitly to the electron scattering as ${\rm Im}\Sigma_{1}({\bf k},\omega)\approx  2\pi[\bar{\Delta}_{\rm PG}({\bf k})]^{2}\delta(\omega+\varepsilon_{0{\bf k}}) $, also reflecting a fact that the product of $[\bar{\Delta}_{\rm PG}({\bf k}_{\rm F})]^{2}$ and the delta function $\delta(\varepsilon_{0{\bf k}_{\rm F}})$ has the same momentum dependence on EFS as that of $|{\rm Im}\Sigma_{1}({\bf k}_{\rm F})|$. To show the momentum dependence of the pseudogap on EFS clearly, we plot the angular dependence of the $|{\rm Im}\Sigma_{1}({\bf k},0)|$ on EFS at $\delta=0.15$ with $T=0.002J$ in Fig. \ref{pseudogap}, where $|{\rm Im}\Sigma_{1}({\bf k},0)|$ [then the pseudogap $\bar{\Delta}_{\rm PG}({\bf k}_{\rm F})$] is strongly anisotropic in momentum space. The most striking feature is that the actual minimum of $|{\rm Im}\Sigma_{1}({\bf k},0)|$ [then the pseudogap $\bar{\Delta}_{\rm PG} ({\bf k}_{\rm F})$] does not appear around the node, but locates exactly at the hot spot ${\bf k}_{\rm HS}$, and then the charge-order state is driven by the Fermi-arc instability, with a characteristic wave vector corresponding to the hot spots of EFS \cite{Comin15,Comin14,Comin15a,Campi15,Harrison14,Atkinson15,Feng16}. This pseudogap opening around the node is also consistent with the experimental observations \cite{Sakai13,Kapon16}. On the other hand, the magnitude of $|{\rm Im}\Sigma_{1}({\bf k},0)|$ [then the pseudogap $\bar{\Delta}_{\rm PG} ({\bf k}_{\rm F})$] still exhibits the largest value around the antinode, and then it decreases with the move of the momentum away from the antinode. In particular, the magnitude of $|{\rm Im}\Sigma_{1}({\bf k},0)|$ [then the pseudogap $\bar{\Delta}_{\rm PG}({\bf k}_{\rm F})$] around the node is smaller than that around the antinode. This special momentum dependence of $|{\rm Im}\Sigma_{1}({\bf k},0)|$ [then the pseudogap $\bar{\Delta}_{\rm PG} ({\bf k}_{\rm F})$] therefore suppresses heavily the low-energy spectral weight of $B_{\rm 1g}$ spectrum, but has a more modest effect on ERR in the $B_{\rm 2g}$ channel.

In conclusion, within the framework of the kinetic-energy-driven SC mechanism, we have studied pseudogap effect on the ERR spectrum in cuprate superconductors in the SC-state by taking into account the interplay between the pseudogap and SC gap. Our results show that the low-energy $B_{\rm 2g}$ spectrum depends linearly on energy $\omega$, while the low-energy $B_{\rm 1g}$ spectrum displays a cube energy dependence. In particular, the pseudogap is strong momentum dependent, where the magnitude of the pseudogap around the nodes is smaller than that around the antinodes. This special structure of the pseudogap therefore leads to that the low-energy spectral weight in the $B_{\rm 1g}$ orientation is suppressed heavily by the pseudogap, while the pseudogap has a more modest effect on ERR in the $B_{\rm 2g}$ channel. Our theoretical results are in qualitative agreement with the experimental data.

\section*{Acknowledgements}

The authors would like to thank Deheng Gao and Yinping Mou for helpful discussions. PJ, YL, LK, and SF are supported by the National Key Research and Development Program of China under Grant No. 2016YFA0300304, and National Natural Science Foundation of China (NSFC) under Grant No. 11574032, and HZ is supported by NSFC under Grant No. 11547034.

\end{document}